\documentstyle[epsfig,twocolumn,aps,amssymb,graphics]{revtex}

\begin{document}
\draft

\def\br{{\bf r}}
\def\bp{{\bf p}}
\def\bv{{\bf v}}

\title{Accidental suppression of Landau damping of the transverse breathing
 mode in elongated Bose-Einstein condensates}
\author{B. Jackson and E. Zaremba}
\address{Department of Physics, Queen's University, Kingston, Ontario 
 K7L 3N6, Canada.}
\date{\today}
\maketitle
\begin{abstract}
 We study transverse radial oscillations of an elongated Bose-Einstein 
 condensate using finite temperature simulations, in the context of a recent
 experiment at ENS. We demonstrate the existence of a mode corresponding to
 an in-phase collective oscillation of both the condensate and
 thermal cloud. Excitation of this mode accounts for the very small 
 damping rate observed experimentally, and we find excellent quantitative 
 agreement between experiment and theory. In contrast to other condensate
 modes, interatomic collisions are found to be the dominant damping
 mechanism in this case. 
\end{abstract}

The study of collective excitations provides us with a powerful tool for 
probing the fundamental properties of quantum many-body systems. For example,
measurements of low-lying excitations in a Bose-Einstein condensed 
(BEC) gas of trapped atoms \cite{jin96} helped establish
the Gross-Pitaevskii equation (GPE) as an excellent description of condensate
dynamics at low temperatures \cite{stringari96}. At higher temperatures the 
situation is complicated by the existence of a significant noncondensed
fraction. Interactions with this thermal cloud leads to
damping and frequency shifts of condensate collective modes, as measured
in several experiments \cite{jin97,stamperkurn98,marago01,chevy02}. A
theoretical understanding of these processes is important as a foundation
for continuing studies in this area, as well as providing a fascinating 
opportunity for studying quantum systems at finite temperatures. 

One can think of interatomic scattering processes as being comprised of mean
fields and collisions, where their relative importance in describing 
the condensate dynamics normally depends upon the rate of collisions 
compared to the characteristic timescale of the oscillation. When 
collisions are relatively frequent, the coupled dynamics of
the condensate and thermal cloud are most appropriately described by
hydrodynamic Landau-Khalatnikov two-fluid equations \cite{nikuni01}. 
It is well known that in liquid helium this leads to hydrodynamic first and 
second sound modes. In the 
opposite, collisionless regime, mean field interactions become more important.
With the possible exception of 
Ref.\ \cite{stamperkurn98}, experiments have so far resided in this 
regime. Here, coupling between fluctuations in the condensate and normal
thermal cloud densities leads to Landau damping and an associated frequency
shift of the condensate mode, and has been the subject of several theoretical
treatments \cite{liu97,giorgini00}. However, the thermal cloud can play
a role beyond that of simply providing a damping mechanism. It 
can undergo collective motion of its own, so 
that the response of the system to an external probe can give rise to a
coupled oscillation of both the condensate and thermal cloud. This kind of
mean field coupling is responsible \cite{bijlsma99,jackson02a} for some of the
behavior observed in the experiment of Ref.\ \cite{jin97}.   

In this paper we study the transverse breathing mode of a 
cigar-shaped condensate at finite temperatures, as recently observed in 
an experiment at ENS \cite{chevy02}. As we shall see, the coupled dynamics of
both the condensate and thermal cloud are particularly important in this 
situation, giving rise to somewhat surprising behavior. Such
considerations are crucial in providing an understanding of the very small 
damping rates and frequency shifts measured in this experiment.

Consider a Bose gas with atoms of mass $m$ confined in an axisymmetic
harmonic trap potential of the form $V (\br)=m[\omega_{\bot}^2(x^2+y^2)+
\omega_z^2 
z^2]/2$. The ratio between the axial and radial trap frequencies,
$\lambda=\omega_z/\omega_{\bot}$, characterizes the trap anisotropy. In this
work we will apply the formalism of \cite{zaremba99}, where the dynamics
of the system are described
by a generalized GPE for the condensate wavefunction, $\Phi (\br, t)$,
\begin{equation}
 i\hbar \frac{\partial \Phi}{\partial t} = \left(
 -\frac{\hbar^2}{2m} \nabla^2 + V + g [n_c + 2 \tilde{n}]-iR \right) 
 \Phi,
\label{eq:GP-gen}
\end{equation}
and, within a semiclassical Hartree-Fock approximation,
a Boltzmann equation for the thermal cloud phase-space distribution,
$f(\bp,\br,t)$,
\begin{equation} 
 \frac{\partial f}{\partial t} + \frac{{\mathbf{p}}}{m} \cdot \nabla f
 - \nabla U_{\rm eff} \cdot \nabla_{\mathbf{p}} f 
 = C_{12} [f] + C_{22} [f].
\label{eq:kinetic}
\end{equation}
The condensate and thermal cloud densities are given by $n_c (\br,t) = 
|\Phi (\br,t)|^2$ and $\tilde{n} (\br,t) = \int ({\rm d}\bp/h^3)
f(\bp,\br,t)$, respectively. In the above equations, an important quantity is
the interaction parameter $g=4\pi \hbar^2 a/m$ (where $a$ is the $s$-wave
scattering length). It not only enters into the effective mean field
potential $U_{\rm eff}=V+2g(n_c + \tilde{n})$, but also into the collision
integrals $C_{22}$ (which represents binary collisions involving only 
thermal atoms) and $C_{12}$ (involving a condensate atom in the incoming
or outgoing channels). The effect of atom exchange in the $C_{12}$ collision 
process appears in (\ref{eq:GP-gen}) through the non-Hermitian source term
$R(\br,t)=(\hbar/2n_c)\int ({\rm d}\bp/h^3) C_{12} [f]$.

First of all, it is instructive to summarize the behavior of the system in 
the $T=0$ limit and for temperatures above the transition, $T>T_c$. The 
former corresponds to a pure condensate, described by (\ref{eq:GP-gen}) with 
$\tilde{n}=0$ and $R=0$. For an axisymmetric anisotropic trap 
($\lambda \neq 1$), analysis of the GPE 
in the interaction-dominated Thomas-Fermi (TF) limit gives two
low-energy $m=0$ modes with frequencies 
$\omega_{\pm}^2/\omega_{\bot}^2=2+3\lambda^2/2 \pm (9\lambda^4-16
 \lambda^2+16)^{1/2} /2$ \cite{stringari96}.
The high-lying mode $\omega_+$ will be of
interest here, which for a cigar-shaped trap with $\lambda \ll 1$
corresponds to a breathing-type oscillation in the radial direction, 
with a frequency $\omega_+ \rightarrow 2 \omega_{\bot}$ as 
$\lambda \rightarrow 0$. In this limit the radial oscillations are decoupled
from those in the axial direction. Interestingly, for a non-interacting
condensate the mode frequency is also $2\omega_{\bot}$ for all $\lambda$, so 
that the transverse mode frequency in the $\lambda \rightarrow 0$ limit is 
independent 
of interactions. This behavior is similar to that found theoretically for a 
gas in a 2D harmonic trap interacting with a contact potential 
\cite{kagan96}, where an underlying hidden symmetry gives rise to 
an undamped breathing mode with frequency $\omega=2 \omega_{\bot}$,
independent of interaction strength, temperature, or statistics. For the
ENS experiment, $\lambda \simeq 6.46\times 10^{-2}$, and the high-lying mode 
has the TF frequency $\omega \simeq 2.00052\, \omega_{\bot}$, in good 
agreement with the experimentally measured value. 

Above $T_c$, the thermal cloud evolves according to
(\ref{eq:kinetic}) with $n_c=0$ and $C_{12}=0$. The mode frequencies can
then be calculated by taking moments of the kinetic equation
\cite{gueryodelin99}. In the hydrodynamic limit the transverse mode 
frequency is $\omega_{\rm HD} = \sqrt{10/3} \omega_{\bot}$ 
\cite{gueryodelin99,griffin97} for $\lambda \rightarrow 0$, while the 
collisionless result is 
$\omega_{\rm CL} = 2 \omega_{\bot}$. The ENS experiment resides in the 
near-collisionless regime, where the mode has frequency $\omega \simeq 
\omega_{\rm CL}$ and is weakly damped through interatomic collisions. 
So, one has the interesting situation where both the condensate at $T=0$
and the thermal cloud at $T>T_c$ have mode frequencies close to 
$2\omega_{\bot}$. 

The question then arises as to the behavior for  
$0<T<T_c$, where both the condensate and thermal cloud are 
present. Answering
this question requires a self-consistent solution of the coupled GPE 
(\ref{eq:GP-gen}) and Boltzmann equations (\ref{eq:kinetic}). We perform this
task numerically using a 
split-operator/Monte-Carlo simulation, as described in \cite{jackson02b}. 
As we shall show, a mode exists at finite temperatures corresponding to an
in-phase motion of the two components with a frequency of $\omega \simeq 2 
\omega_{\bot}$, which is a direct consequence of the behavior in the $T=0$ 
and $T>T_c$ limits. This mode, which is only weakly damped, corresponds to a
breathing oscillation of the entire trapped gas. We demonstrate that
this mode was predominately excited in the ENS experiment \cite{chevy02}, 
leading to the anomalously small measured damping rate and frequency shift. 

We first discuss results of our simulations for a temperature of 
$T=125\, {\rm nK}$, where all other parameters are chosen to match the 
experiment \cite{chevy02,footnote1}. At this temperature, which should be 
compared to the experimentally measured critical temperature of 
$T_c \simeq 290\, {\rm nK}$, we find a condensate fraction of 0.56. 
We calculate the radial moments, $\langle x^2 + y^2 \rangle = 
\int {\rm d}\br\, n(x^2+y^2)$, as a function of time for both the 
condensate and thermal clouds. This moment projects out the transverse
breathing mode, and is related to the condensate squared radius measured in 
Ref.\ \cite{chevy02}. 

Three distinct excitation schemes are considered. The first 
matches that employed experimentally, where the radial trap 
frequency was abruptly changed, then reset to its original
value after a time $\tau$. This procedure can be represented by 
$\omega_{\bot}' (t) = \omega_{\bot}\{1 + \alpha [\Theta (t) - \Theta 
(t-\tau)] \}$,
where $\Theta (x)$ is the Heaviside step function. In Fig.\ 1(a) we show 
results for $\alpha=0.26$ and $\omega_{\bot} \tau= 0.172$. The figure clearly
shows that both components respond by oscillating with approximately equal 
amplitudes and phase, at a frequency very close to $2 \omega_{\bot}$. 
The in-phase oscillation of the condensate and thermal cloud continues over
the entire timescale of the simulation, with both amplitudes decaying
at a very similar rate. Importantly, this damping is very weak, much 
smaller than one would expect
for Landau damping which is usually the dominant damping mechanism in this 
regime.

\begin{figure}
\centering
\psfig{file=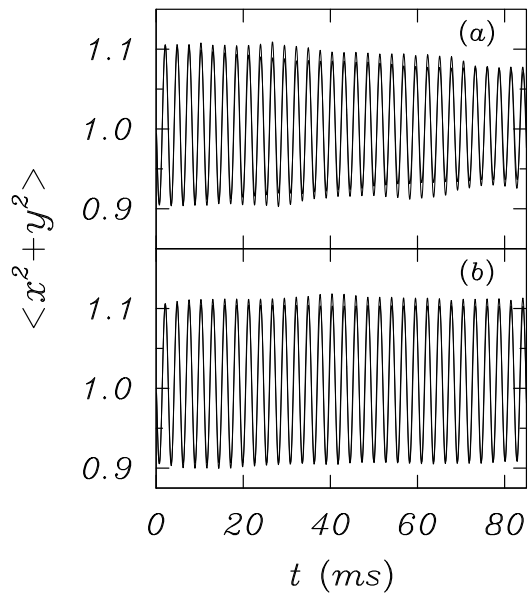, scale=0.38, bbllx=40, bblly=75, 
 bburx=560, bbury=680} 
\caption{\label{excboth}
 Time-dependent moment $\langle x^2+y^2 \rangle$ for the condensate
 (narrow line) and thermal cloud (bold line), divided by the corresponding 
 values at $t=0$. Data is for a temperature of $T=125\,{\rm nK}$ and is the
 result of exciting the system using the tophat perturbation employed 
 experimentally {\protect \cite{chevy02}}.
 Simulations shown in (a) include collisions, while those in (b) exclude all 
 collisions ($C_{12}=C_{22}=0$).}
\end{figure}

To understand this behavior one should note that calculations of Landau 
damping \cite{liu97,giorgini00} conventionally
assume that the noncondensate is always in thermal equilibrium. 
Fluctuations of the condensate mean field induce a response in the
thermal cloud density, where the dissipative and reactive components 
in turn give rise to damping and frequency shifts of the condensate mode. 
However, this picture is clearly inadequate when the thermal 
cloud is itself in oscillation. In this case, mean field coupling
may lead to the thermal cloud driving the condensate at its own resonant
frequency, which can be significant when the natural frequencies of the 
two components
are nearly degenerate. As shown in \cite{jackson02a}, a similar 
near-resonant driving of the condensate is in fact responsible for the abrupt
change in mode frequency observed at high temperatures in Ref.\ \cite{jin97}. 
The present case can be seen as 
an extreme example of this behavior, as the natural frequencies of the 
condensate
and thermal cloud are both very close to $2\omega_{\bot}$. Mean field
interactions then lead to phase-locking of the oscillations of the two 
components.   

We test this interpretation by separately exciting the two components in our
simulations. This is readily achieved by imposing 
velocity fields at $t=0$ of the
form $\bv \propto x \hat{\bf i} + y \hat{\bf j}$ \cite{jackson02a,footnote2}
on either the condensate or the thermal cloud. Fig.\ 2(a) shows the result
of exciting the condensate only. The condensate oscillation is initially 
damped quite strongly, while the oscillation in the thermal cloud slowly
builds up in time. Eventually, the
condensate and thermal cloud settle down to a small amplitude in-phase mode.
The initial condensate relaxation is essentially Landau damping due to 
mean field interaction with the stationary thermal cloud, with a damping rate
similar to that found for the $m=2$ mode discussed later.  
This excitation scheme could be implemented experimentally by 
optically imprinting a phase onto the condensate, and would be useful in 
verifying the physical picture presented here. 

\begin{figure}
\centering
\psfig{file=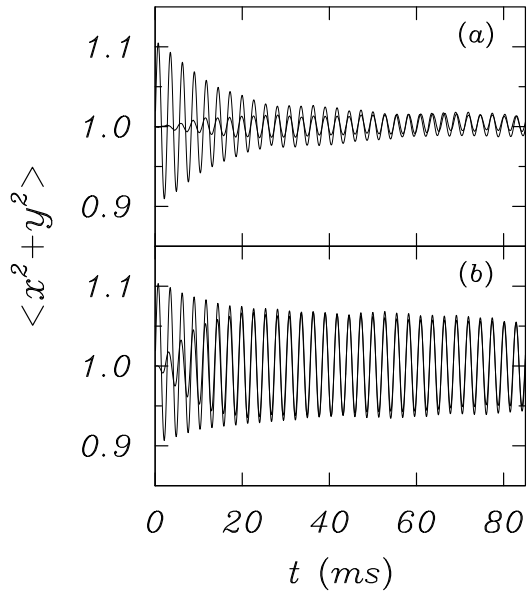, scale=0.38, bbllx=40, bblly=75, 
 bburx=560, bbury=680}
\caption{\label{excone}
 Results of initially exciting only the (a) condensate, and (b) thermal cloud
 by imposing velocity fields on each, for a temperature $T=125\,{\rm nK}$.
 In (a) the condensate moment is represented by the narrow line with the 
 thermal 
 cloud in bold, while in (b) the line styles are interchanged for clarity.} 
\end{figure}

In Fig.\ 2(b) we show the result of exciting the thermal cloud only. Again, 
we see the relaxation of one component (thermal cloud) matched by excitation 
of the other (condensate), with both settling into an in-phase oscillation at
later times. The larger amplitude
of the in-phase mode in this case is a consequence of the much greater spatial
extent of the thermal cloud. On imposing the 
same velocity field, the energy increase is proportional to the
equilibrium value of $\langle x^2+y^2 \rangle$, which is several 
times larger for the thermal cloud than for the condensate at the temperature
considered. During the 
subsequent evolution, the initial energy of either component is redistributed
until the two components oscillate together in phase. Since the energy of 
this in-phase mode 
is mostly stored in the thermal cloud, one can see that the final 
amplitude should indeed be larger when the thermal cloud is excited initially.

To compare our results directly to experiment, we fit the condensate data 
for our first excitation scheme [as displayed in Fig.\ 1(a)] to an
exponentially decaying sinusoid $A \cos(\omega t + \varphi) 
{\rm e}^{-\Gamma t} + C$. Our results for the frequency $\omega$, and
damping rate $\Gamma$ \cite{footnote3} are plotted in Fig.\ 3 for a range of 
temperatures,
together with the corresponding experimental data. We find very good 
agreement between experiment and theory, for both the frequency and
damping rate \cite{footnote4}. It is interesting to note that, along with the
very small
damping, the frequency is virtually independent of temperature in both 
simulations and experiment. This is in marked contrast to experimental
\cite{jin97,stamperkurn98,marago01} and theoretical results 
\cite{liu97,giorgini00,jackson02a} for other modes, and is a unique feature 
of this breathing mode excitation. We note that fits to the thermal cloud 
data give results very similar to those in Fig.\ 3.

So far we have not addressed the origins of the damping. To provide insight 
into this question, we have performed a simulation at $T=125\, {\rm nK}$ where
collisions are artificially switched off. Thus,  
the only coupling between the two components is provided by the mean field
terms. The result is plotted in Fig.\ 1(b). The most striking feature of 
this plot, when compared to Fig.\ 1(a), is the almost complete absence of 
damping, either in the condensate
or the thermal cloud. Specifically, a fit to the condensate moment
yields a damping rate an order
of magnitude smaller than the collisional result. Furthermore, the 
separate inclusion
of either $C_{12}$ or $C_{22}$ collisions yields values 
intermediate between
the collisionless and full collisional results. Finally, artificially 
increasing the collision cross-section by a factor of two leads to an 
increase in the damping rate by a corresponding factor. These simulations, 
together with the excellent agreement found with experiment, provide strong
evidence that collisions play a crucial role in determining the observed
damping. 

We should emphasize that the supression of Landau damping
discussed here is specific to the $m=0$ transverse breathing mode in a 
highly prolate harmonic trap, and is a consequence of the accidental
degeneracy between the condensate and thermal cloud frequencies. This 
condition is no longer satisfied, for example, by the lowest $m=2$
mode, since the condensate now has a frequency of approximately 
$\sqrt{2}\omega_{\bot}$ \cite{stringari96} while the thermal cloud frequency 
remains close to $2\omega_{\bot}$ \cite{gueryodelin99}. A simulation at
$T=125\,{\rm nK}$ for this mode indeed shows a damping rate of 
$\Gamma \simeq 50\, {\rm s}^{-1}$, an order of magnitude larger than that of
the $m=0$ mode and in agreement with the qualitative result mentioned in 
\cite{chevy02}. Moreover, we would also
expect larger damping rates for the $m=0$ mode as $\lambda \rightarrow 1$,
since the condensate TF frequency approaches $\sqrt{5} \omega_{\bot}$ 
in this limit. Simulations at $\lambda=0.57$ and $\lambda=0.75$ confirm
that lifting the degeneracy between the condensate and thermal cloud modes 
leads to an increase in damping with decreasing trap anisotropy. 

\begin{figure}
\centering
\psfig{file=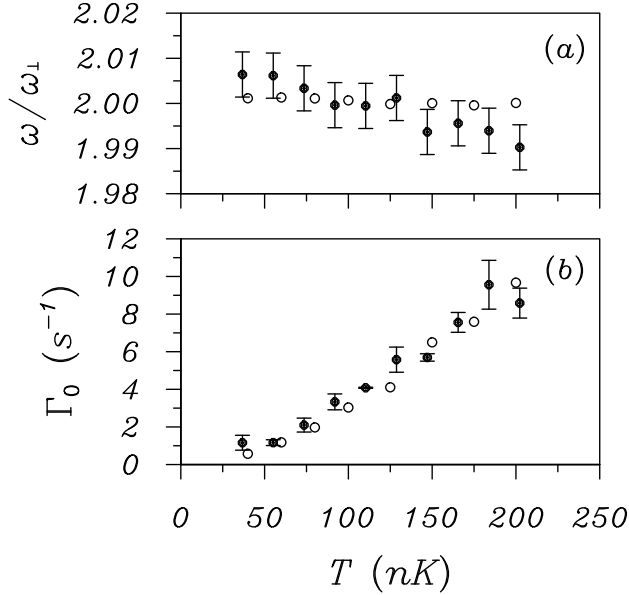, scale=0.43, bbllx=25, bblly=100, 
 bburx=580, bbury=650}
\caption{\label{freqdamp} Frequency (a) and damping rate (b) of the 
 condensate breathing mode. Our results (open circles) are compared to the 
 experimental data (solid circles), where the simulation parameters and 
 excitation scheme are chosen to reproduce the experimental conditions
 {\protect \cite{chevy02}}.} 
\end{figure}

In summary, we have used finite-temperature simulations to study the
transverse breathing mode in elongated Bose-Einstein condensates, as
realized in a recent experiment \cite{chevy02}. We show that the very small
damping rate and frequency shift observed experimentally is a result of a 
degeneracy between the condensate and thermal cloud oscillation frequencies,
which results in a mode comprised of an in-phase oscillation of both 
components. The weak damping arises through interatomic binary collisions, 
rather than mean field interactions (Landau damping) as is normally the case.
A comparison of simulation results to experiment shows excellent agreement, 
confirming the validity of our theoretical approach and underlining the 
importance of
treating the full noncondensate dynamics, including collisions, when 
accounting for experimental behavior.  

We thank V. Bretin for useful information and for providing us with 
the experimental 
data in Fig.\ 3. We acknowledge use of the HPCVL computing facility at
Queen's 
University, and financial support from Natural Sciences and Engineering
Research Council of Canada.


\begin{references}
\bibitem{footnote1} In particular, we consider $^{87}{\rm Rb}$ atoms confined
 with a radial trap frequency of 
 $\omega_{\bot} = 2 \pi \times 182.6\, {\rm Hz}$. Experimental use of
 evaporative cooling meant that the total number of atoms decreased with 
 temperature. We account for this variation in our simulations.
\bibitem{footnote2} Note that in the limit $\omega_{\bot} {\tau} \ll 1$
 the tophat excitation employed experimentally corresponds to a delta
 function impulsive perturbation. In linear response this is 
 equivalent to imposing identical velocity fields of the form given on both 
 the condensate and thermal cloud.
\bibitem{footnote3} Due to small departures of the condensate data from an 
 exponentially-decaying form, our determination of the damping rate has 
 some uncertainty, which we estimate to be about $15\,\%$.
\bibitem{footnote4} Experimentally a condensate amplitude revival was 
 observed at low temperatures and after strong excitations. We see no 
 evidence of this behavior in simulations that extend over periods twice as 
 long as 
 those in Fig.\ 1. Possible sources for this effect are discussed in
 \cite{chevy02}.
\end{references}
\end{document}